\documentclass[a4paper,11pt]{article}
\usepackage{jcappub}
\usepackage{amssymb}
\usepackage{amsmath}
\usepackage{lineno}
\usepackage[T1]{fontenc}
\usepackage{graphicx}
\usepackage{tikz}

\def\be            {\begin{equation}}
\def\bearl         {\begin{array}{l}}
\def\bearll        {\begin{array}{ll}}
\def\ee            {\end{equation}}

\newcommand \Mdm  {M_{\scriptscriptstyle{DM}}}

\title{Estimates for a DM$\rightarrow a \rightarrow \gamma$ 3.55\,keV line in the radio lobes of Centaurus A}

\author{Nicholas Jennings}

\affiliation{Rudolf Peierls Centre for Theoretical Physics, Oxford}

\emailAdd{nicholas.jennings@physics.ox.ac.uk}

\abstract{The origin of the 3.55\,keV excess reported in various galaxies and galaxy clusters remains unknown. Non-observations of the line in dwarf spheroidal galaxies (dSphs) create tension with galaxy cluster observations when considering a direct Dark Matter (DM)~$\rightarrow \gamma$ decay. This discrepancy could be explained by a DM~$\rightarrow a \rightarrow \gamma$ decay, where the DM particles decay to axion-like particles (ALPs), which then convert to photons in a magnetic field. The 3.55\,keV line from a decay to ALPs therefore has very different morphology from a direct decay to photons, as it would only appear in environments with substantial magnetic fields between the source and us. To test this hypothesis we consider targets where the line strength would be enhanced compared to the DM~$\rightarrow \gamma$ model. The giant lobes of radio galaxies could represent good environments, having magnetic fields with strength $\mathcal{O}(\mu\rm{G})$ that extend for hundreds of kiloparsecs, and minimal X-ray background. In the case of Centaurus~A, a large population of dSphs are associated with the galaxy, some of which are potentially in or behind its lobes. In this paper we estimate the flux of a 3.55\,keV DM~$\rightarrow a \rightarrow \gamma$ line from these sources, and compare it to the flux from the DM halo of Centaurus~A. We comment on the potential to detect such a signal with current satellites (such as {\it XMM-Newton} and {\it Chandra}) and future satellites (such as {\it Athena}).}

\begin{document}
\maketitle
\flushbottom
\providecommand{\e}[1]{\ensuremath{\times 10^{#1}}}

\section{Introduction}

The claimed detections of an unidentified emission line at 3.55\,keV in stacked samples of galaxy clusters and individual galaxies \cite{Bulbul2014DETECTIONCLUSTERS,Boyarsky2014UnidentifiedCluster.} have generated a huge amount of interest among particle physicists. While the possibility remains that this could be an astrophysical effect, no single theory has been able to account for all features of the data. This has led to many models of a Dark Matter (DM) particle decaying to photons to explain the line (see \cite{Iakubovskyi:2015wma} for a review). However, the differing strengths of the line in different systems are in tension with a direct decay to photons.\\
\\
This discrepancy could be resolved by a DM $\rightarrow a \rightarrow \gamma$ decay, where an intermediate ALP particle then converts to photons in the presence of a magnetic field. Such a decay would explain many features of the data \cite{Conlon:2014xsa,Conlon:2014wna}, including: why the strength of the line is bounded to be weak for dwarf spheroidal galaxies (dSph), where there are no substantial magnetic fields \cite{Malyshev:2014xqa,Ruchayskiy:2015onc,Jeltema:2015mee}; and why it is measured to be strongest in the Perseus cluster, where the presence of an extended $\mathcal{O}(\mu\rm{G})$ magnetic field in the intracluster medium has been established \cite{Alvarez:2014gua}. It therefore behoves us to consider targets where the morphology for a DM~$\rightarrow a \rightarrow \gamma$ decay differs substantially from direct DM~$\rightarrow \gamma$.\\
\\
The giant lobes of radio galaxies represent a promising environment to test the DM~$\rightarrow a \rightarrow \gamma$ model. These lobes can extend for hundreds of kiloparsecs and contain magnetic fields $\mathcal{O}(\mu\rm{G})$ and low electron densities ($\lesssim\,10^{-4}\,\rm{cm}^{-3}$), similar to galaxy clusters that have been shown to be efficient ALP-photon converters \cite{Conlon:2015uwa}. Objects such as dSphs in or behind the lobes, with large DM to baryonic matter ratios, might produce a 3.55\,keV line competitive with the small X-ray background. In contrast, dSphs not along the l.o.s. to to the lobes will have no associated 3.55\,keV.\\
\\
Centaurus A is the best radio galaxy candidate to search for a line as it is the closest, with 500\,kpc long lobes and an associated population of dSphs \cite{Karachentsev2007TheComplex}. There is also some evidence that the magnetic field is very strong ($13.4\,\mu\rm{G})$ in a region of the southern lobe far removed from the host galaxy, while other regions have magnetic fields $\sim 1\,\mu\rm{G}$ \cite{Sun:2016ibh}. This could allow for a direct test of the DM$\rightarrow a \rightarrow \gamma$ model, as the signal from a dSph here should be much stronger than for a dSph in another part of the lobe. It would also be interesting to determine the strength of a signal from the DM halo of Centaurus A and how this compares to dSphs.\\
\\
We review the current status of the 3.55\,keV line in Section~\ref{3.5kev}, and discuss potential radio galaxy targets in Section~\ref{radio}. In Section~\ref{conversion}, we review ALP-photon conversion in magnetic fields, and describe the magnetic field model used to derive conversion probabilities. We compare 3.55\,keV signatures for dSphs behind the lobes and the DM halo of Centaurus A in Section~\ref{calculation}, and conclude in Section~\ref{conclusion}.

\begin{table}
\centering
\begin{tabular}{|c|c|c|c|}
\hline
Sample & Instrument & Energy & $\sin^2(2\theta) \times 10^{-11}$\\
\hline
Perseus & XMM-MOS & $3.57$ &$23.3^{+7.6}_{-8.9}$\\
 & XMM-PN & &$< 18~(90\%)$\\ 
 & Chandra ACIS-I & $3.56 \pm 0.02$ &$28.3^{+11.8}_{-12.1}$ \\
& Chandra ACIS-S & 3.56 &$40.1^{+14.5}_{-13.7}$\\ \hline
Coma + Centaurus + Ophiuchus & XMM-MOS & 3.57 &$18.2^{+4.4}_{-5.9}$\\
& XMM-PN & &$< 11~(90\%)$\\ \hline
69 stacked clusters (Bulbul) & XMM-MOS & 3.57 &$6.0^{+1.1}_{-1.4}$ \\
& XMM-PN & 3.57 &$5.4^{+0.8}_{-1.3}$ \\ \hline
M31 on-centre & XMM-MOS & $3.53 \pm 0.03$ &2--20 \\ \hline
Stacked galaxies & XMM-Newton & &$< 2.5~(99\%)$ \\ \hline
Stacked galaxies & Chandra & &$< 5~(99\%)$ \\ \hline
Stacked dwarves & XMM-Newton & &$< 4~(95\%)$ \\ \hline
Draco & XMM-Newton & &$\lesssim 2 - 5~(95\%)$ \\ \hline
\end{tabular}
\caption{The inferred line strength in different systems, observed with different instruments. References for the values can be found in Section~\ref{3.5kev}.}
\label{linestrengths}
\end{table}

\section{Review of the 3.55\,keV line}
\label{3.5kev}

The 3.55\,keV line was initially detected in 2014 in a stacked sample of 73 galaxy clusters and in the Perseus cluster \cite{Bulbul2014DETECTIONCLUSTERS}, and by a separate group in M31 and the Perseus cluster \cite{Boyarsky2014UnidentifiedCluster.}. These studies involved a total of 4 separate detectors on two different satellites (ACIS-I and ACIS-S onboard {\it Chandra}, and MOS and PN onboard {\it XMM-Newton}), making it difficult to explain as a systematic effect. A potential New Physics (NP) explanation proposed for the 3.55\,keV excess was a 7.1\,keV DM sterile neutrino $\psi$ that decays to a 3.55\,keV photon and neutrino. The decay rate is given by:

\be
\Gamma_{\psi \rightarrow \gamma\nu} = \frac{9\alpha_{EM}}{1024\pi^4}\sin^2(2\theta)G_F^2m_\psi^5.
\ee
\\
Since then, analyses have disagreed on the existence of the line in the centre of the Milky Way \cite{Jeltema:2014qfa,Boyarsky:2014ska,Riemer-Sorensen:2014yda} and in {\it Suzaku} observations of the Perseus cluster \cite{Urban:2014yda,Tamura:2014mta,Franse:2016dln}. Observations of dwarf spheroidals \cite{Malyshev:2014xqa,Jeltema:2015mee} have not found evidence for a 3.55\,keV line. The different inferred decay rates from the various (non-)observations are in tension (although not excluded). They are summarised in Table~\ref{linestrengths}, parametrised in terms of $\sin^2(2\theta)$.\\
\\
Several astrophysical explanations of the 3.55\,keV line have been proposed. Of these, the most compelling have been: an emission line caused by higher than expected abundances of ionised potassium (K XVIII) at 3.51\,keV \cite{Jeltema:2014qfa}; or sulphur charge exchange at 3.44--3.47\,keV, where highly ionised gas interacts with cold neutral gas, causing transitions from high n orbitals to the ground state \cite{Gu:2015gqm,Shah:2016efh}. The latter explanation would require an almost identical gain miscalibration across 4 different detectors, in the same direction, to explain why the line appears at 3.55\,keV.\\
\\
In 2016 the {\it Hitomi} satellite observed the Perseus cluster for 230\,ks, and found no evidence for a 3.55\,keV line \cite{Aharonian:2016gzq}. This observation ruled out the potassium explanation, and was in $3\sigma$ disagreement with the higher inferred line strength from {\it XMM-Newton's} MOS instrument (it was, however, consistent with line strengths inferred from other instruments and sources). Unfortunately, with the loss of {\it Hitomi}, confirmation that the line has a New Physics origin will have to wait either for the {\it Hitomi} replacement X-ray Astronomy Recovery Mission {\it XARM} or, looking further ahead, the {\it Athena} X-ray observatory due to launch in 2028 \citep{Nandra:2013shg}.\\
\\
The variation in the inferred line strengths, and the lack of a definitive astrophysics explanation, motivates consideration of alternative NP models to explain the line. One possibility is that there is no overall emission of 3.55\,keV photons, but instead an absorption and rapid re-emission of 3.55\,keV photons from point sources that were excluded from the above analyses. This would point to a 2-state Fluorescent Dark Matter model \cite{Berg:2016ese,Conlon:2016lxl}. Another model that explains the data well is a decay from DM to an intermediate particle, whose conversion to photons depends on the astrophysical environment between the DM halo and us. If the intermediate particle is an axion-like particle (ALP), the probability of conversion to photons is greatly enhanced in the presence of large astrophysical magnetic fields \cite{Raffelt1988MixingParticles}, producing a substantially different morphology for the line \cite{Cicoli:2014bfa}, as the decay rate:

\begin{equation}
\label{decayrate}
\Gamma_{\rm{DM} \rightarrow a \rightarrow \gamma}({\bf B}) = \Gamma_{\rm{DM} \rightarrow a} P_{a \rightarrow \gamma}({\bf B})
\end{equation}
\\
depends on the magnetic field ${\bf B}$ through the ALP-photon conversion term $P_{a \rightarrow \gamma}$.\\
\\
The dependence on ${\bf B}$ could explain the variety of inferred values for $\sin^2(2\theta)$ in Table~\ref{linestrengths}. The Perseus cluster is known to host a large magnetic field that efficiently converts ALPs to photons, so it would be likely to produce a strong signal \cite{Berg:2016ese}. An extremely weak signal would be expected from dSphs, as these small objects cannot host a magnetic field that could cause substantial ALP-photon conversion \cite{Beck:2013bxa,Spekkens:2013ik}. The predictions for the Milky Way are less clear due to the uncertainties in the structure of the magnetic field towards the centre. In the case that the magnetic field is $\mathcal{O}(10-100\,\mu\rm{G})$, no 3.55\,keV line would be expected; in the case of a 1\,mG poloidal central magnetic field, a line could be produced \cite{Alvarez:2014gua}. However, the model does predict a stronger line in M31 than the Milky Way \cite{Conlon:2014xsa}, consistent with the positive detection of \cite{Boyarsky2014UnidentifiedCluster.}.\\
\\
The unique morphology of a DM~$\rightarrow a \rightarrow \gamma$ line would allow future observations to differentiate it from a DM~$\rightarrow \gamma$ scenario. Predicted line strengths differ for cool-core and non-cool-core clusters \cite{Conlon:2014wna}. Stacked samples of nearly edge-on galaxies could provide good targets, as the distance the ALP travels through the galactic magnetic field is maximised \cite{Alvarez:2014gua}. Recently, it has been proposed that a DM~$\rightarrow a \rightarrow \gamma$ line could be detected in upcoming polarisation satellite experiments, such as IXPE \cite{Gong:2016zsb}, albeit with a very long exposure time. The focus of this paper will be the potential to detect a 3.55\,keV line in the giant lobes of radio galaxies. These provide good environments for ALP-photon conversion, and therefore merit a detailed consideration to determine the expected flux from DM halos in or behind the lobes. 

\section{Candidate Radio Galaxies}
\label{radio}

We consider nearby elliptical galaxies hosting AGNs with 10-100\,kpc scale jets, and residing in poor groups or in cluster outskirts. These jets can produce giant radio lobes expanding for up to 500\,kpc, which contain a tangled relic magnetic field. For most radio galaxies, the strength and coherence lengths of the magnetic fields are not known. NGC 6251 and DA240 have lobes with magnetic field strengths estimated \mbox{$\mathcal{O}(0.1 - 1\,\mu \rm{G})$} \cite{Sambruna2004TheNGC6251,Takeuchi2012SUZAKUJ1629.4+8236,Isobe2011Suzaku240}, however we have no knowledge of their coherence lengths.\\
\\
The nearest radio galaxy to us, Centaurus A, represents the best candidate to search for a DM~$\rightarrow a \rightarrow \gamma$ line. Its outer lobes are around 500\,kpc long, and 200\,kpc wide. The coherence length of the magnetic field are limited to be $\gtrsim 10\,\rm{kpc}$ by \cite{OSullivan2009StochasticGalaxies} and $\gtrsim 30\,\rm{kpc}$ by \cite{Wykes2013MassCentaurusA}). The electron number density is limited to be $n_{e} \leq 7 \times 10^{-5}\,$cm$^{-3}$ by \cite{Wykes2013MassCentaurusA}, while \cite{Feain2009FARADAYA} limit it to $n_{e} \leq 5 \times 10^{-5}\,$cm$^{-3}$. While the AGN and jets will produce significant 0.2-10 keV emission, the lobes produce very little. In an analysis of two patches of the southern lobe using the Hard X-ray detector on Suzaku (XIS) \cite{Koyama2007X-RaySuzaku}, the absorbed X-Ray energy flux in the 2-10 keV range was found to be  $F = 6.5 \times 10^{-12}$ erg cm$^{-2}$/0.35deg$^{2}$, a value consistent with the lobe emission being no higher than 10\% of the CXB \cite{Stawarz2013GIANTSATELLITE}.\\
\\
Estimates of its magnetic field strength are typically  $\sim 1\,\mu\rm{G}$, derived from: Fermi-LAT detection of gamma rays interpreted as inverse Compton scattering of CMB photons ($0.89\,\mu\rm{G}$ in the northern lobe, $0.85\,\mu\rm{G}$ in the southern lobe \cite{Abdo2010FermiGalaxy.b}); synchrotron emission ($1\,\mu\rm{G}$ \cite{Hardcastle2013SynchrotronDistributions}) and equipartition arguments ($1.3\,\mu\rm{G}$ \cite{Hardcastle2009High-energyA}). More recently, Fermi-LAT data was used to infer magnetic field strengths of $1\,\mu\rm{G}$ for most of the radio lobes, apart from a region in the southern lobe furthest from the galaxy, where the best-fit value was $13.4\,\mu\rm{G}$ \cite{Sun:2016ibh}. It remains to be seen whether this high value will be corroborated by other studies, and the subtleties of disentangling electron distributions from magnetic field strengths mean this should be treated with caution. Possible contributions from hadronic processes could push the value down to $1\,\mu\rm{G}$. We will therefore consider magnetic field strengths of $1\,\mu\rm{G}$ and $13.4\,\mu\rm{G}$ in our analysis, and stress that the higher value may not be robust\footnote{The jets likely have much stronger magnetic fields associated with them but over shorter coherence lengths. In addition there is much more background X-ray emission. The complexities of modelling such magnetic fields are beyond the scope of this paper, therefore we do not consider conversion in regions near the jets.}.\\
\\
Crucially, Centaurus A has a number of galaxies, including identified dSphs, in its vicinity \cite{Karachentsev2007TheComplex}. Currently there are 40 confirmed dSphs \cite{1997AJ....114.1313C,2000AJ....119..609V,2014ApJ...795L..35C,2016ApJ...823...19C,2015ApJ...802L..25T}, some of which lie along the line of sight to the radio lobes, which extend between approximately $-38^{\circ}$ and $-48^{\circ}$ Declination and between 13h40m and 13h20m Right Ascension \cite{Keivani2015MagneticA} (a list of objects that could lie within the radio lobes based on the Catalog of Neighboring Galaxies are shown in Table~\ref{dsph} \cite{Jerjen2000SurfaceGroups,Karachentsev2004AGalaxies}). Inferred distances indicate some could lie in or behind the lobes, providing a potential source for a 3.55\,keV ALP line. A more recent survey with the {\it Dark Energy Camera} \cite{2015AJ....150..150F} has uncovered potentially 41 new dSph candidates \cite{2015A&A...583A..79M,2016A&A...595A.119M,2017A&A...597A...7M}. Dwarf Spheroidals make ideal sources of DM decay processes due to their high mass-to-luminosity ratios and negligible X-ray emission. In addition, the velocity broadening of a 3.55\,keV line from a dSph is far below 1\,eV, making it easier to discriminate the line from the background than in galaxy clusters \cite{Walker:2007ju}. However, currently there is no information on the DM profiles of these objects. We therefore estimate a signal from a dSph based on DM profiles for the classical dSphs.\\
\\

\begin{table}[h]
\centering
\begin{tabular}{| l | l | l | l | l |}
\hline
\textbf{Object} & \textbf{Type} & \textbf{R.A.} & \textbf{Dec.} & \textbf{Distance(Mpc)}\\
\hline
Cen A &   & 13 25 28.9 & -43 01 00 & $3.77 \pm 0.38$ \\
KK 196 & dIrr & 13 21 47.1 & -45 03 48 & $3.98 \pm 0.29$ \\
KK 197 & dSph & 13 22 01.8 & -42 32 08 & $3.87 \pm 0.27$ \\
KKs 55 & dSph & 13 22 12.4 & -42 43 51 & $3.94 \pm 0.27$ \\
KK 203 & ? & 13 27 28.1 & -45 21 09 & 3.8 \\
E324-24 & lsb & 13 27 37.4 & -41 28 50 & $3.73 \pm 0.43$ \\ 
E270-17 & SBm & 13 34 47.3 & -45 32 51 & $4.3 \pm 0.8$ \\

\hline
\end{tabular}
\caption{Coordinates and distances of objects near the lobes of Centaurus A}
\label{dsph}
\end{table}

\section{Modelling ALP-photon conversion in giant radio lobes}
\label{conversion}

\subsection{Review of ALP-photon conversion}

An ALP couples to electromagnetism through the Lagrangian term:

\begin{equation}
\mathcal{L} \supset \frac{1}{8M} aF_{\mu\nu}\tilde{F}^{\mu\nu} \equiv \frac{1}{M}a\vec{E}\cdot\vec{B}
\end{equation}
where $M^{-1} = g_{a\gamma\gamma}$ is the ALP-photon coupling. For a homogeneous magnetic field domain of length $L$, the probability of ALP-photon conversion is \cite{Raffelt1988MixingParticles,Sikivie:1983ip}:

%


\begin{equation}
P(a \to \gamma) = \sin^2(2\theta)\sin^2\bigg(\frac{\Delta}{\cos(2\theta)}\bigg)
\end{equation}
where $\tan(2\theta) = \frac{2B_\perp\omega}{Mm_{eff}^2}, \Delta = \frac{m_{eff}^2L}{4\omega}$, for energy $\omega$ and magnetic field perpendicular to the ALP wave vector $B_\perp$, and $m_{eff}^2 = |m_a^2 - \omega_{pl}^2|$ for an ALP mass $m_a$ and plasma frequency $\omega_{pl} = \sqrt{4\pi\alpha n_e/m_e}$. Henceforth we assume $m_a \ll \omega_{pl}$ and set it to zero. After plugging in constants, $\tan(2\theta)$ and $\Delta$ evaluate to:

\begin{equation}
\label{theta}
\tan(2\theta) = 4.9 \times 10^{-2} \bigg( \frac{10^{-4} \rm{cm}^{-3}}{n_e} \bigg) \bigg( \frac{B_\perp}{1 \mu \mathrm{G}} \bigg)\bigg( \frac{\omega}{3.5 \mathrm{keV}} \bigg)\bigg( \frac{10^{13} \mathrm{GeV}}{M} \bigg)
\end{equation}
\begin{equation}
\label{Delta}
\Delta = 1.5 \times 10^{-2}\bigg( \frac{n_e}{10^{-4} \rm{cm}^{-3}} \bigg)\bigg( \frac{3.5 \mathrm{keV}}{\omega} \bigg)\bigg( \frac{L}{10\,\mathrm{kpc}} \bigg)
\end{equation}
\\
For $\Delta \ll 1$ and $\theta \ll 1$ the conversion probability simplifies to:

\begin{equation}
\label{Pag}
P(a \to \gamma) = 2.3 \times 10^{-8}\bigg(\frac{B_\perp}{1 \mu \mathrm{G}}\frac{L}{1 \mathrm{kpc}}\frac{10^{13}\mathrm{GeV}}{M}\bigg)^2
\end{equation}
\\
For $M \gtrsim 10^{13}\,\rm{GeV}$, this condition holds in radio lobes, as well as galaxy clusters. The inferred value of $\Gamma_{\rm{DM} \rightarrow a}$ from equation~\ref{decayrate} therefore is proportional to $M^2$. Our calculations for the 3.55\,keV line strength in radio galaxies is therefore independent of $M$.

\subsection{Magnetic field model}
\label{magnetic}

We model the magnetic field of the radio lobes far from the jet region as a multi-scale, random-domain tangled field which was used to model synchrotron emission from Centaurus A in \cite{Hardcastle2013SynchrotronDistributions}. These models have also been used for magnetic fields of galaxy clusters \cite{Murgia2004MagneticGalaxies,Angus2014SoftBackground}. We generate a random vector potential with a power spectrum:

\begin{equation}
\langle |\tilde{A}_k|^2\rangle \sim |k|^{-n}
\end{equation}
\\
The magnitude and phase are uniform for each domain. The magnitude is randomly selected from a Rayleigh distribution:

\begin{equation}
p(\tilde{A}_k) = \frac{\tilde{A}_k}{|k|^{-n}} \exp{\bigg(-\frac{\tilde{A}_k^2}{2|k|^{-n}}\bigg)}
\end{equation}
\\
while the phase is uniformly distributed from 0 to $2\pi$. The one-dimensional power spectrum of the magnetic field is then:
\begin{equation}
\mathcal{P}(k) \sim 2\pi k^2 |\tilde{B}_k|^2 \propto k^{-n+4}
\end{equation}
for $\tilde{B}_k = ik \times \tilde{A}_k$. The value of $n$ is inferred from synchrotron data \cite{Hardcastle2013SynchrotronDistributions}, which supports a value close to Kolmogorov ($n = 17/3$). We model the magnetic field along a 200\,kpc line of sight, which is the width of the radio lobes. We truncate the power spectrum, $k_{min} < k < k_{max}$ where $k_{min} = 2\pi/\Lambda_{max}$ and $k_{max} = 2\pi/\Lambda_{min}$. The minimum length scale $\Lambda_{min} = 10\,\rm{kpc}$ uses the value derived in \cite{OSullivan2009StochasticGalaxies}, while we examine the effect of allowing $\Lambda_{max}$ to vary. We conservatively take the electron number density $n_e = 10^{-4}\,\rm{cm}^{-3}$ to be constant throughout the lobe. We therefore also take the magnetic field strength to be constant across 200\,kpc, and 0 outside the lobe. We model both $1\,\mu\rm{G}$ and $13.4\,\mu\rm{G}$.\\
\\
We generated 1000 different magnetic field configurations for each field strength. We propagated a 3.55\,keV ALP from cell-to-cell and calculated the total conversion probability for each configuration. In all cases we used $M = 10^{13}\,\rm{GeV}$. For a 20 cell model with 10\,kpc length domains (i.e. $\Lambda_{max} = \Lambda_{min} = 10\,\rm{kpc}$), an average conversion probability of $(2.5 \pm 0.1) \times 10^{-5}$ was derived for a magnetic field strength of $1\,\mu\rm{G}$, and $(4.4 \pm 0.1) \times 10^{-3}$ for a magnetic field strength of $13.4\,\mu\rm{G}$. Here we quote value $\pm$ standard deviation based on the variation of magnetic field configurations: this is not a full account of the error. If we take $\Lambda_{max} = 30\,\rm{kpc}$, over 200\,kpc the conversion probabilities are enhanced by no more than a factor of 2. The conversion probability for $1\,\mu\rm{G}$ is 2 orders of magnitude lower than that typical for galaxy clusters, while for $13.4\,\mu\rm{G}$ it is of the same order.


\section{Estimating X-ray signals}
\label{calculation}

\subsection{Signal from dark matter in a dSph}

The brightnesses of the dSphs near Centaurus A are not well constrained, making an estimation of their DM profiles challenging \cite{Karachentsev2002NewGroup}. We therefore estimate DM profiles based on those of the classical dSphs. The astrophysical D-factors of the classical and ultrafaint dSphs have been calculated in \cite{Bonnivard2015DarkDSphs}. As the strongest signal will come from the centre of the dSph, we wish to use the best-fit Einasto profiles calculated in this paper rather than just the D-factor, where the Einasto density profile is given by:
\begin{equation}
\rho_{Ein}(r) = \rho_{-2}\exp{\Big\{-\frac{2}{\alpha}\Big[\Big( \frac{r}{r\,_{-2}}\Big)^\alpha - 1 \Big]\Big\}}.
\end{equation}
\\
We used the best constrained dSph density profiles (Leo I and II, CVI, Carina, Fornax, Sculptor, Draco, Ursa Minor and Sextans) and integrate along the l.o.s. to produce a 2D integrated DM column density profile. While the DM density profile at the centre is model-dependent, the integrated density profile receives a small contribution from the central volume of the dSph. We found the densities of the central kpc$^{2}$ of the dSphs lie between $10^7 - 10^8$ M$_{\odot}$/kpc$^2$. We use this to estimate the ALP flux from the central kpc$^2$, which corresponds to 1\,arcmin at 3.8\,Mpc. The decay rate from the DM particle to ALPs is inferred from Perseus observations to be:

\begin{equation}
\Gamma_{\rm{DM} \rightarrow a} \sim 2 \times 10^{-25}\bigg(\frac{M}{10^{13} \, \mathrm{GeV}}\bigg)^2 \,\rm{s}^{-1}
\end{equation}
\\
where the dependence on $M$ compensates for the fact that $P_{a \rightarrow \gamma} \propto 1/M^2$. Using \mbox{$M=10^{13}\,\rm{GeV}$}, and the values of $P_{a \rightarrow \gamma}$ calculated in Section~\ref{magnetic}, we calculate the total 3.55\,keV flux:

\be\label{DM_flux1}
F_{{\scriptscriptstyle{DM}} \rightarrow a \rightarrow \gamma}=\frac{\Gamma_{\scriptscriptstyle{DM}\rightarrow a}}{4\pi \, d^2} P_{a\rightarrow \gamma}\int \rho_{\scriptscriptstyle{DM}} \, \mathrm{d}V
\ee
\\
where the DM density $\rho_{DM}$ is integrated over a volume equal to 1\,kpc$^2 \times l$ where $(4/3)\pi(l/2)^3$ is large enough to include more than 99\% of the DM mass, and $d$ is the distance to the dSph.
From a typical dSph, the flux is found to be $1 - 10 \times 10^{-20}$\,erg\,s$^{-1}$\,cm$^{-2}/\rm{arcmin}^2$ in the case of a $1\,\mu\rm{G}$ magnetic field strength and $1 - 10 \times 10^{-18}$\,erg\,s$^{-1}$\,cm$^{-2}/\rm{arcmin}^2$ for $13.4\,\mu\rm{G}$. The X-ray background between 2-10 keV is $1.2 \times 10^{-14}$ erg s$^{-1}$ cm$^{-2}/\rm{arcmin}^2$. Therefore we estimate that for a 100\,eV detector resolution (such as for the instruments onboard {\it Chandra} and {\it XMM-Newton}) a background flux of $\sim 1.5 \times 10^{-16}$\,erg\,s$^{-1}$\,cm$^{-2}/\rm{arcmin}^2$, and for a 2.5\,eV detector resolution (such as for the X-ray Integral Field Unit onboard {\it Athena} \cite{Barret:2016ett}) a background flux of $\sim 4 \times 10^{-18}$\,erg\,s$^{-1}$\,cm$^{-2}/\rm{arcmin}^2$.\\
\\
In the case that we have a dSph behind a region of the radio lobe where the magnetic field strength is $1\,\mu\rm{G}$, it would be very challenging to detect a 3.55\,keV line, as we would need to be sensitive to a 1\% effect with a satellite like {\it Athena}. This would require substantial satellite time and also precise modelling of the contribution from X-ray emission from the radio lobe. In the case of a magnetic field strength of $13.4\,\mu\rm{G}$, the situation is more promising. With {\it Athena}, the 3.55\,keV line would be an effect of the same order as the CXB, which could be detectable. However, a $13.4\,\mu\rm{G}$ magnetic field strength is anomalously high compared to the rest of the radio lobe, and the data may be explained by other sources of emission not taken into account by the model. Were this value to be confirmed, it would be worth observing dSphs within this region to search for a 3.55\,keV line. If the magnetic field strength is found to be $\sim 1\,\mu\rm{G}$ everywhere in the lobe, our analysis shows that dSphs are not promising targets for a 3.55\,keV DM~$\rightarrow a \rightarrow \gamma$ line.

\subsection{Signal from the Dark Matter halo of Centaurus A}
\label{dmcena}

It is instructive to compare the signals we would expect from dSphs to that produced by the DM halo of Centaurus A itself. In the case of a DM$\rightarrow \gamma$ decay the line strength should be substantially greater in Centaurus A as it hosts a much larger DM halo. However, with a $1\,\mu\rm{G}$ magnetic field strengths in the regions of the lobes near the galaxy, we might expect the DM$\rightarrow a \rightarrow \gamma$ line to be weaker.\\
\\
To estimate the DM profile of Centaurus A we follow the procedure used in \cite{Anderson2015Non-detectionSpectra}. In order to calculate an NFW profile:
\be
\rho_{\scriptscriptstyle{NFW}}=\frac{\rho_0}{\frac{r}{r_s}(\frac{r}{r_s}+1)^2}\,\,,
\ee
\\
we infer the parameters $\rho_0$ and $r_s$ from the K-band apparent magnitudes listed in the 2MASS All-Sky Extended Source Catalog. From the K-band magnitudes ($m_{\scriptscriptstyle{K}}$) we infer the stellar masses $m_s$, setting:
\be\label{mL}
\frac{m_s}{m_{\odot}}\frac{L_{\odot,\scriptscriptstyle{K}}}{L_{\scriptscriptstyle{K}}}=0.5\,.
\ee
\\
We then determine the total DM mass $\Mdm$ within the virial radius using eq. 13 in \cite{Moster2010CONSTRAINTSREDSHIFT}:
\be
m_s=2 \Mdm \left(\frac{m_s}{\Mdm}\right)_0  \left[ \left(\frac{\Mdm}{M_{1}} \right)^{-\beta} +  \left(\frac{\Mdm}{M_{1}} \right)^{\gamma}\right]\,,
\ee
\\
where $\left(\frac{m_s}{\Mdm}\right)_0=0.0282$,   $\beta=1.06$, $\gamma=0.556$, $\text{log}M_1=11.884$. We estimate the virial radius as:
\be
R_{\text{vir}}=\left( \frac{\Mdm}{200 \frac{4}{3}\pi \rho_c} \right)\,,
\ee
\\
with $\rho_c=9.1\cdot\,10^{-30}\, \text{g\, cm}^{-3}$. Following \cite{Prada2012HaloCosmology}, we compute the halo concentration $c=c(\Mdm, z)$.\\

\begin{figure}[h]
\centering
\includegraphics[scale=0.375]{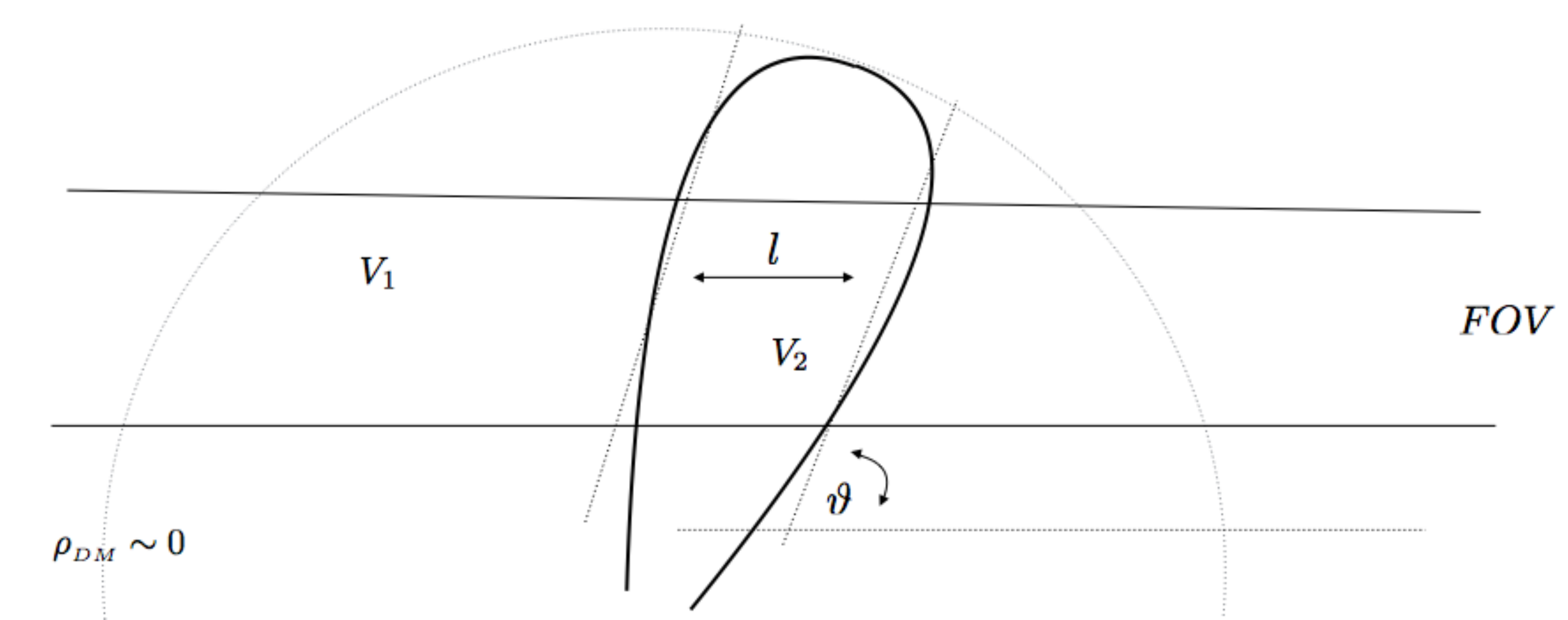}
      \caption{Diagram of the regions $V_1$ and $V_2$ as defined in Equation~\ref{DM_flux2}. The observer is to the right of the diagram, with $V_1$ being the volume of the Centaurus A DM halo behind the lobe, and $V_2$ the volume within the lobe.}
        \label{lobe}
\end{figure}
\begin{figure}[h]
\centering
\begin{tikzpicture}
\node[above right] (img) at (0,0) {\includegraphics[scale=0.35]{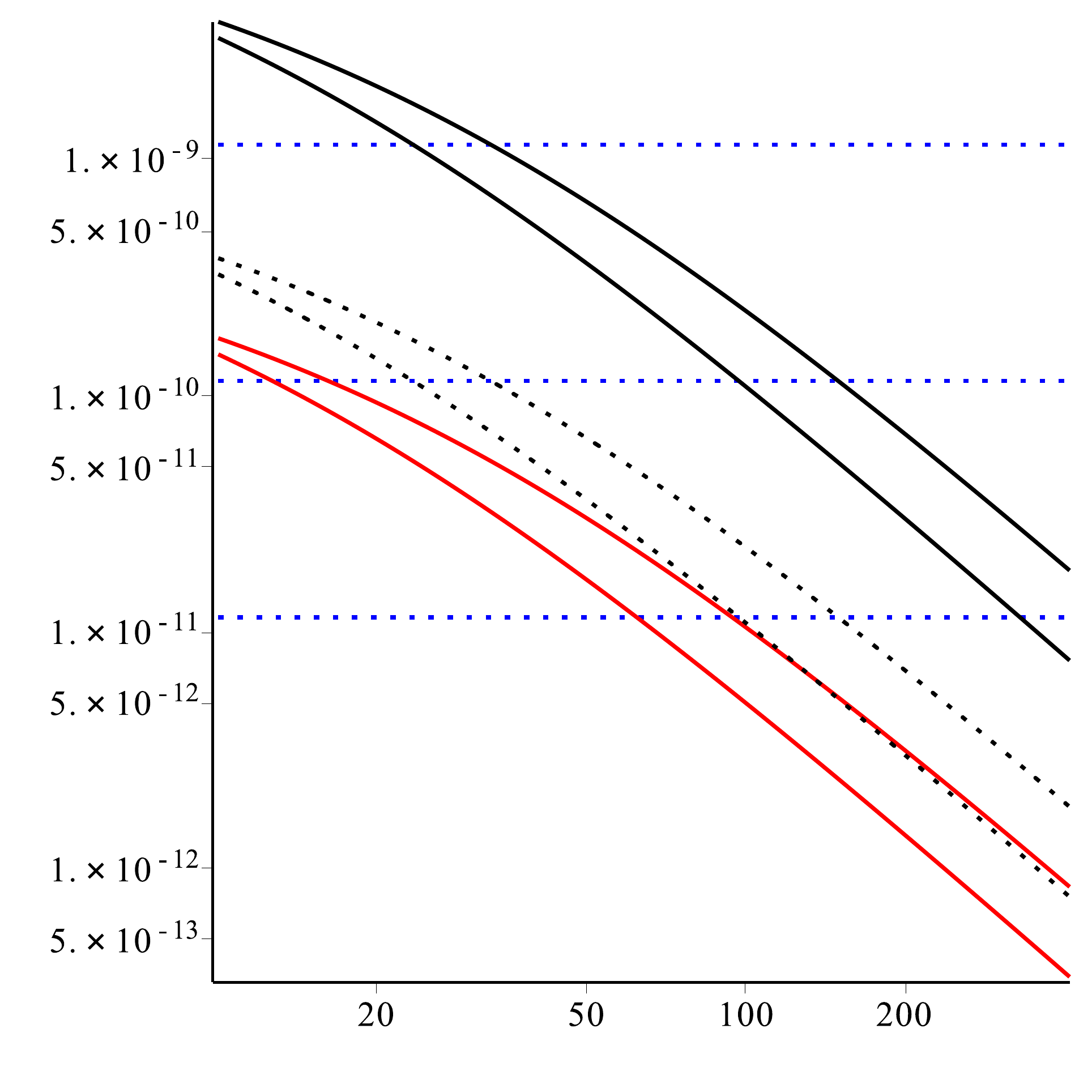}};
\node at (0,6.9) {$S \left (\frac{\text{cnts}\,\, \text{s}^{-1}}{\text{arcmin}^{2} \text{cm}^{2} }\right)$};
\node at (6.6,0.2) {$r\,$(Kpc)};

\node at (5.1, 5.4) {$\scriptstyle{DM\rightarrow \gamma}$};
\node at (2.8, 3.5) {$\scriptstyle{DM\rightarrow a \rightarrow \gamma}$};

\node at (7.5, 6.2) {$\scriptstyle{10\%}$};
\node at (7.5,4.7) {$\scriptstyle{1\%}$};
\node at (7.5,3.2) {$\scriptstyle{0.1\%}$};
\end{tikzpicture}
        \caption{The 3.55\,keV line surface brightness as a function of the distance from the core of Centaurus~A. The solid black line corresponds to a $DM\rightarrow \gamma$ direct decay, \mbox{$\Gamma_{\scriptscriptstyle{DM}\rightarrow \gamma}=2 \times 10^{-28}\,\rm{s}^{-1}$}, computed with two different NFW profiles (see Section~\ref{dmcena}) while for the dotted black line \mbox{$\Gamma_{\scriptscriptstyle{DM}\rightarrow \gamma}=2 \times 10^{-29}\,\rm{s}^{-1}$}. The solid red line corresponds to a $DM\rightarrow a \rightarrow \gamma$ decay. The dotted blue lines show the brightness as a percentage of the background emission within a 100\,eV range.}
        \label{CentA}
\end{figure}

Finally we compute $\rho_0$ and $r_s$ from
\be
c=\frac{R_{\text{vir} } }{r_s}
\ee
and 
\be
\Mdm=4\pi \int_0^{R_{\text{vir}}} \rho_{\scriptscriptstyle{NFW}}\, r^2 \, dr\,.
\ee
\\
For Centaurus~A we find that $\rho_0 = 0.003\,(M_{\odot} /\text{pc}^3)$ and $r_s = 29.3\,\rm{kpc}$. Alternative derivations from direct kinematic measurements and an isothermal fit produce similar results \cite{Peng2004The5128}. The three profiles are compared in Figure~\ref{CentA_nfw}.\\
\\
In this case, the 3.55\,keV flux calculation is subtly different to Equation~\ref{DM_flux1}, as we must consider the ALP flux from the DM halo behind the lobe separately from the DM halo within the lobe:

\begin{figure}[h]
\centering
\begin{tikzpicture}
\node[above right] (img) at (0,0) {\includegraphics[scale=0.35]{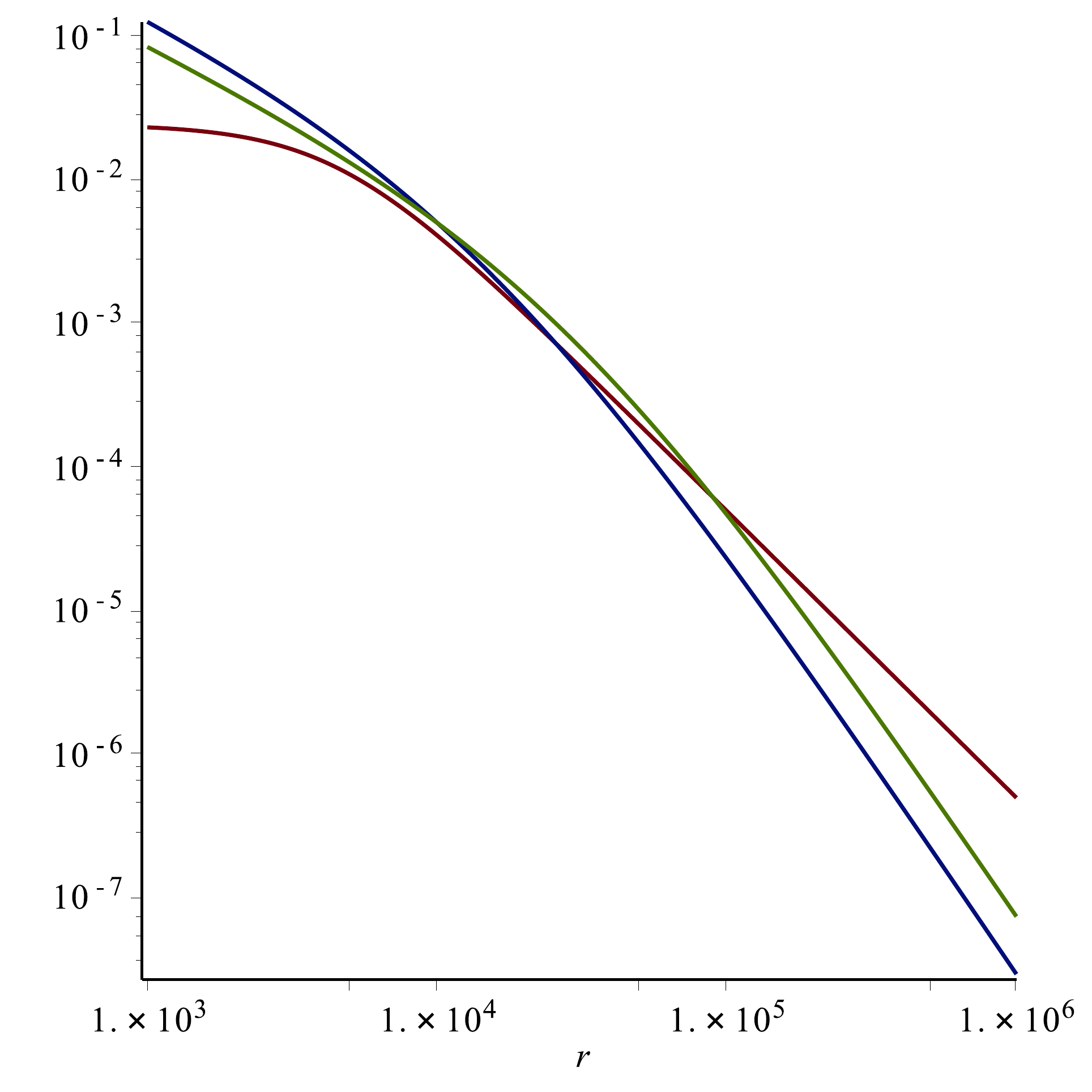}};
\node at (0,6.5) {$\rho$};
\node at (6.5,0) {$r\,$(pc)};
\end{tikzpicture}
        \caption{Comparison of Centaurus~A dark matter profiles.  Green represents the NFW profile derived from the K-band magnitude \cite{Anderson2015Non-detectionSpectra};  blue, NFW fitted from kinematic measurements in \cite{Peng2004The5128}; purple, isothermal fitted in \cite{Peng2004The5128}.}
        \label{CentA_nfw}
\end{figure}

\be
F_{{\scriptscriptstyle{DM}} \rightarrow a \rightarrow \gamma}\sim\frac{\Gamma_{\scriptscriptstyle{DM}\rightarrow a}}{4\pi \, d^2}  \left( 
 P_{a\rightarrow \gamma}^l \int_{V_1} \rho_{\scriptscriptstyle{DM}}\, \,dV
 +
\int_{V_2} \rho_{\scriptscriptstyle{DM}}\, P_{a\rightarrow \gamma}(V) \,dV
\right)
\label{DM_flux2}
\ee
\\
where $P_{a\rightarrow \gamma}^l$ is the ALP-photon conversion probability across the propagation length $l$, $P_{a\rightarrow \gamma}(V)$ the ALP-photon conversion probability across a volume $V$, $V_1$ is the non magnetised volume within the field of view back to the magnetised lobes, and $V_2$ is the lobes volume within the FOV approximated as a cylinder, as shown in Figure~\ref{lobe}. $P_{a \gamma \gamma}$ is calculated using the magnetic field model of Section~\ref{magnetic} with a field strength of $1\,\mu\rm{G}$.\\
\\
The expected 3.55\,keV line flux from the DM halo of Centaurus A is plotted against distance in Figure~\ref{CentA} for two different NFW profiles. It is also plotted against the expected fluxes for a DM$\rightarrow \gamma$ line of the strength observed in the Perseus Cluster, and an order of magnitude weaker as observed in other galaxy clusters. As can be seen, the DM~$\rightarrow a \rightarrow \gamma$ flux is substantially smaller than that of a direct decay to photons, which is over 10\% of the CXB up to 50\,kpc from the centre for the higher value of $\sin^2(2\theta)$. Therefore a non-observation of the 3.55\,keV line in the DM halo of Centaurus A could provide a consistency check on an observation of a 3.55\,keV line in dSphs behind the radio lobes for a DM~$\rightarrow a \rightarrow \gamma$ model.

\section{Conclusion}
\label{conclusion}

The origin of the 3.55\,keV line remains an intriguing open problem. The DM~$\rightarrow a \rightarrow \gamma$ decay model produces a unique morphology for the line that allows for unique testable predictions for different astrophysical environments. It is therefore important to analyse all promising targets to search for a  DM~$\rightarrow a \rightarrow \gamma$ line. The giant radio lobes of Centaurus~A contain a relic magnetic field that extends over hundreds of kiloparsecs, and could have a magnetic field strength up to $13.4\,\mu\rm{G}$. We have discussed the possibility to detect a signal by observing dwarf spheroidal galaxies behind these radio lobes. For a magnetic field strength of $1\, \mu\rm{G}$, and a galaxy of similar dark matter profile to the classical dSphs, the ALP-photon conversion probability is $\mathcal{O}(10^{-5})$, which would produce a signal $\mathcal{O}(10^{-4} - 10^{-3})$ of the CXB within a 100\,eV energy range, and $\mathcal{O}(10^{-2} - 10^{-1})$ for a 2.5\,eV range, meaning that we are unlikely to be able to detect a signal. If the anomalously high magnetic field strength of $13.4\,\mu\rm{G}$ inferred for a region in the southern lobe is confirmed, this could boost the signal to be of the same order as the CXB in a 2.5\,eV range. In this case, it could be possible to detect a 3.55\,keV line from a dSph, which would provide a compelling test of the DM~$\rightarrow a \rightarrow \gamma$ hypothesis. Further observations to better determine the dSph density profiles, and the magnetic field strength of the radio lobes, are needed to determine a DM~$\rightarrow a \rightarrow \gamma$ could be observed.

\bibliographystyle{jhep}
\bibliography{Mendeley.bib}

\end{document}